\documentclass[preprint,aps,showpacs,showkeys,dvips]{revtex4}

\usepackage{graphicx}
\usepackage{dcolumn}
\usepackage{bm}
\usepackage{subfigure}

\begin{document}

\title{Excited Quark Production at Future $\gamma p$ Colliders}

\author{R. \c{C}ift\c{c}i}
\email{rena.ciftci@gmail.com}
\affiliation{Dept. of Eng. of Physics, Faculty of Eng., Ankara University, 06100 Tandogan,
Ankara, Turkey}

\date{\today}

\begin{abstract}
Excited quark production  at future $\gamma p$ colliders is studied. Namely, $\gamma p \rightarrow q^{*}X$ with subsequent  $q^{*}\rightarrow gq$ and $\gamma q $ decay channels are considered. Signatures for discovery of the excited quark and corresponding standard model backgrounds are discussed in detail. Discovery limits for excited quark masses and achievable values of compositeness parameters $f_s$, $f$ and $f^\prime $ are determined.
\end{abstract}
\maketitle
\section{Introduction}	

Composite models are foreseen to explain repeated generations of quarks and leptons. Excited quarks with rich spin and weak isospin spectra are unavoidable outcome of preonic models [1-6]. Therefore, observation of excited quarks and investigation of their properties will give opportunity to determine mechanism of compositeness of quarks and leptons. 

D0 experiment at Fermilab Tevatron collider excludes excited quarks with mass region below 775 GeV for two jet decays at $p \bar{p}$ collisions with the assumption of $f=f'=f_s=\Lambda/m^*$ [7]. Also, mass region below 760 GeV are excluded for $jj$ decay by CDF experiment [8, 9]. Recently, D0 experiment has assumed $q^*$ production via $q g$ fusion and via contact interactions and looked at Z jet decay to obtain exclusion of below 510 GeV [10]. The quoted limit is for $\Lambda = m^*$.
 
The best place to discover the excited quark is hadron colliders like CERN Large Hadron Collider (LHC) [6, 11]. Very high mass reach and low coupling values are possible at LHC. Unfortunately, LHC does not permit accurate investigation of the properties of the excited quarks. Excited quarks can be pairly produced at $e^+e^-$ colliders. However, their production is restricted by relatively low center of mass energy of lepton colliders. At this point of view $ep$ and $\gamma p$ colliders are more advantageous because the excited quark is single produced at them. In addition to clearer signal environment, it is possible to investigate properties of the excited quarks with photo-production [12-14]. In $ep$ colliders, Weisz\"{a}cker-Williams photons are used to produce excited quarks. However, their spectrum is concentrated at lower energies. Fortunately, it is possible to produce real photons by colliding laser photons with high energy electrons. Resulting Compton backscattered photons are produced with high polarization and a spectrum with most of photons carrying about 80 percent of electron energy [15-17]. Recently, LHC based linac-ring type $ep$ colliders were proposed in refs. [18-20] with the names of QCD Explorer with $\sqrt{s}=1.4$ TeV and Energy Frontier $ep$ collider with $\sqrt{s}=3.74$ TeV. In refs. [21-24], the feasibility of $\gamma p$ colliders based on QCD Explorer and Energy Frontier $ep$ colliders is shown.  
 
In this paper, excited quark production  at future $\gamma p$ colliders is studied. In Section II, an effective Lagrangian describing transition between an excited quarks and the standard model (SM) quarks is presented; the decay width and branching ratios of the excited quarks are evaluated. Production of excited quarks at QCD Explorer based $\gamma p$ collider with $\sqrt{s_{max}}=1.27$ TeV center of mass energy and Energy Frontier $\gamma p$ collider with $\sqrt{s_{max}}=3.41$ TeV center of mass energy is studied in Section III: $\gamma p \rightarrow q^{*} X \rightarrow g q X$ and $\gamma p \rightarrow q^{*} X \rightarrow \gamma q X$ processes as well as their SM backgrounds are considered; the statistical significance of the signal and achievable values of compositeness parameters $f_s$, $f$ and $f^\prime $ are evaluated. Finally, concluding remarks are made in Section IV.

\begin{table}
\caption{The total decay width $\Gamma$ of the excited quarks into ordinary quarks and gauge bosons $V=g,W,Z,\gamma$, and branching ratios for $\Lambda=m^{*}$ and $f=f'=f_s=1$.}
\begin{ruledtabular}
\begin{tabular}{ccccccccccc}
$m^{*}$&\multicolumn{4}{c} {$u^{*}$}& $\Gamma_{Tot}$&\multicolumn{4}{c} {$d^{*}$}& $\Gamma_{Tot}$  \\
\cline{2-5}\cline{7-10}
(GeV) & $ug$ &$dW^{+}$ & $uZ^{0}$ & $u \gamma$ &  (GeV)& $dg$ &$uW^{-}$ & $dZ^{0}$ & $d \gamma$ &  (GeV) \\ 
\colrule

600 &  0.8419 &  0.1040 & 0.03209 & 0.02194 &  23.75 & 0.8419 &  0.1040 & 0.04801 & 0.00549 &  23.74\\
700 &  0.8384 &  0.1062 & 0.03287 & 0.02226 &  27.31 & 0.8384 &  0.1062 & 0.04917 & 0.00557 &  27.30\\
800 &  0.8358 &  0.1081 & 0.03349 & 0.02254 &  30.82 & 0.8358 &  0.1081 & 0.05009 & 0.00564 &  30.81\\
900 &  0.8337 &  0.1096 & 0.03401 & 0.02280 &  34.29 & 0.8337 &  0.1096 & 0.05085 & 0.00570 &  34.28\\
1000 & 0.8316 &  0.1110 & 0.03445 & 0.02302 &  37.73 & 0.8316 &  0.1110 & 0.05151 & 0.00576 &  37.72\\
1500 & 0.8243 &  0.1158 & 0.03599 & 0.02389 &  54.54 & 0.8243 &  0.1158 & 0.05382 & 0.00597 &  54.53\\
2000 & 0.8196 &  0.1189 & 0.03699 & 0.02449 &  70.93 & 0.8196 &  0.1189 & 0.05531 & 0.00612 &  70.93\\
2500 & 0.8160 &  0.1213 & 0.03775 & 0.02496 &  86.99 & 0.8160 &  0.1213 & 0.05643 & 0.00624 &  86.99\\
3000 & 0.8130 &  0.1232 & 0.03835 & 0.02535 & 102.81 & 0.8130 &  0.1232 & 0.05734 & 0.00634 & 102.81\\
3500 & 0.8106 &  0.1249 & 0.03886 & 0.02568 & 118.41 & 0.8106 &  0.1249 & 0.05810 & 0.00642 & 118.41\\
4000 & 0.8085 &  0.1263 & 0.03929 & 0.02596 & 133.87 & 0.8085 &  0.1263 & 0.05875 & 0.00649 & 133.87\\
4500 & 0.8066 &  0.1275 & 0.03968 & 0.02621 & 149.17 & 0.8066 &  0.1275 & 0.05932 & 0.00655 & 149.17\\
5000 & 0.8050 &  0.1285 & 0.04000 & 0.02642 & 164.41 & 0.8050 &  0.1285 & 0.05981 & 0.00660 & 164.41\\
\end{tabular}
\end{ruledtabular}
\end{table}

\begin{figure} 
\subfigure[]{\includegraphics[width=8.15cm]{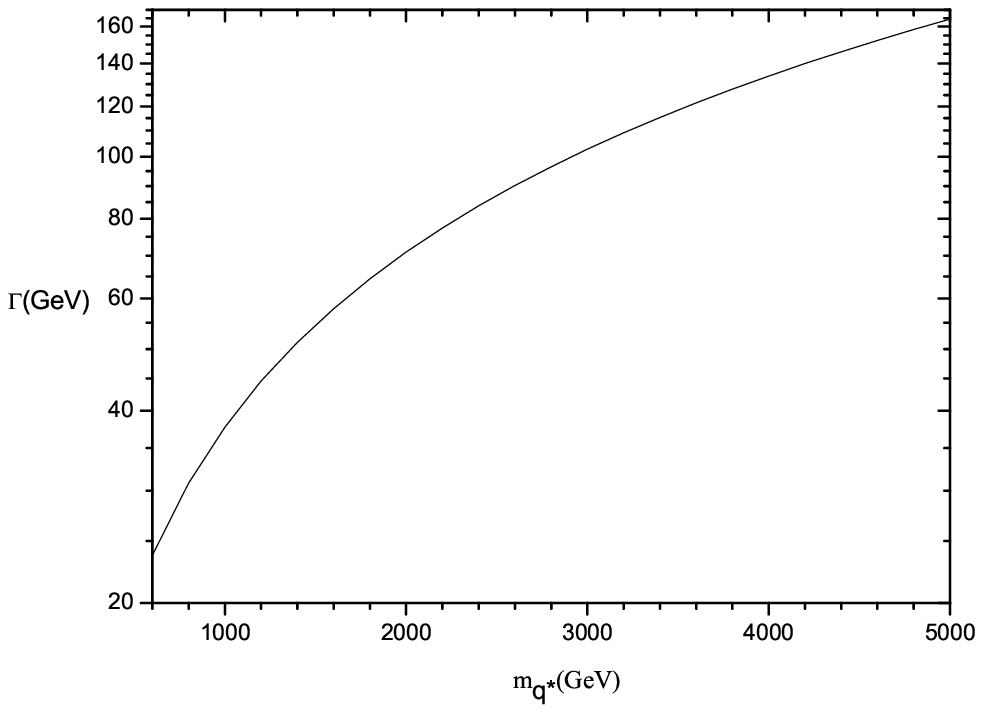}}
\subfigure[]{\includegraphics[width=8.15cm]{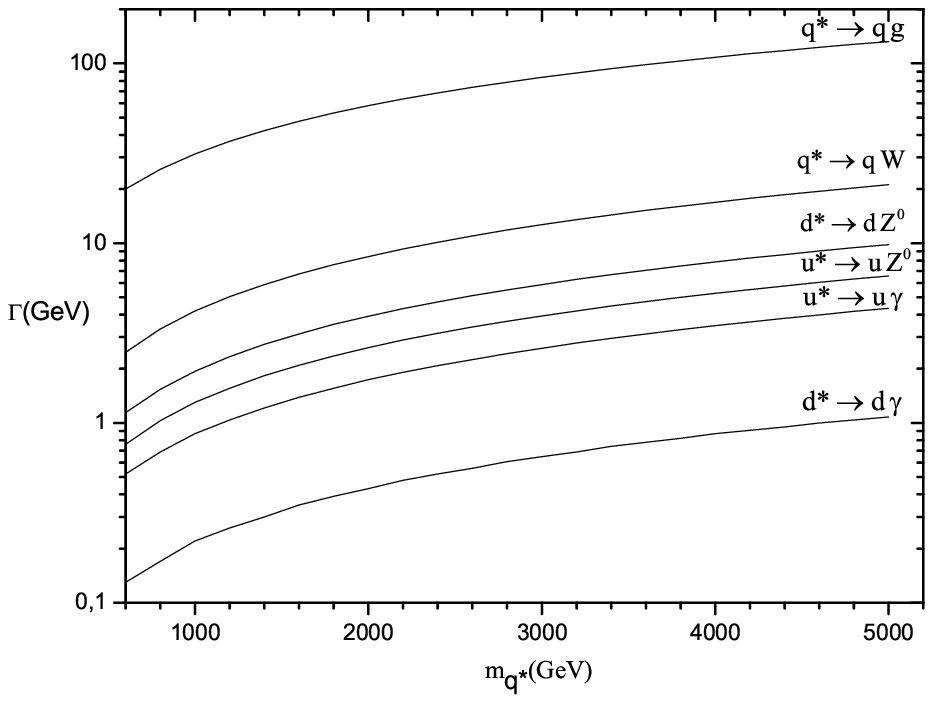}}
\caption{(a) The total and (b) the partial decay widths of the excited quarks as a function of the excited quark mass. \protect\label{fig1}}
\end{figure}

\begin{figure}
\subfigure[]{\includegraphics[width=8.15cm]{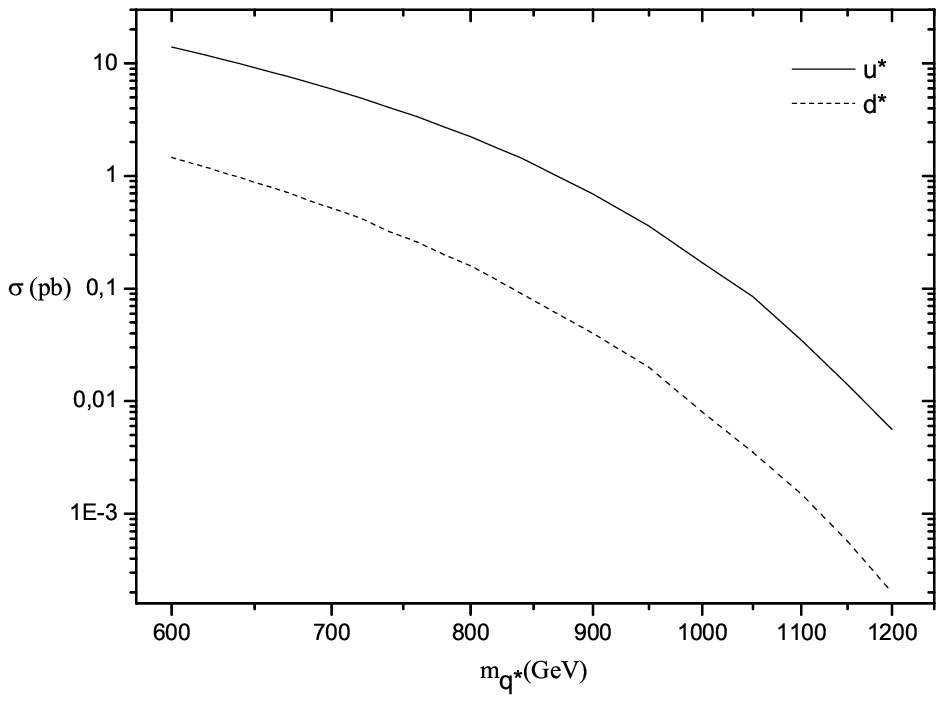}}
\subfigure[]{\includegraphics[width=8.15cm]{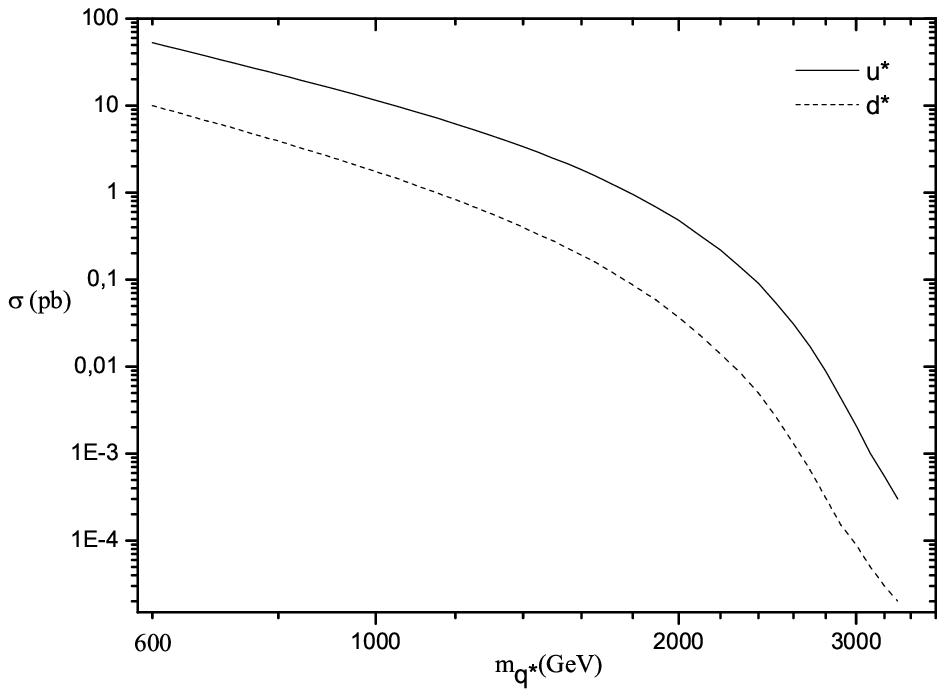}}
\caption{ The total production cross section of the process $\gamma p \rightarrow q^{*} X$ as a function of the excited quark mass with the center of mass energy (a) $\sqrt{s}$ = 1.27 TeV and (b) $\sqrt{s}$ = 3.41 TeV . \protect\label{fig2}}
\end{figure}

\section{Excited Quark Involved Interactions}
 
Excited quarks will have vertices with SM quark and gauge bosons (photon, gluon, Z or W bosons). They can be produced at hadron colliders via quark gluon fusion or at $ep$ and $\gamma p$ colliders via quark photon fusion. Interactions involved excited quark can be described as below [6]:

\begin{equation}
L=\frac{1}{2\Lambda}\bar{q}^*_R \sigma^{\mu\nu}\left(g_s f_s\frac{\lambda^a}{2}G^a_{\mu\nu}+g f \frac{\tau}{2}\cdot W_{\mu\nu}+g'f'\frac{Y}{2} B_{\mu\nu}\right)q_L+h.c.
\end{equation}
\\
where $G^a_{\mu\nu}$, $W_{\mu\nu}$ and $B_{\mu\nu}$ are the field-strength tensors of the gluon, the SU(2) and the U(1) gauge fields; $\lambda^a$, $\tau$ and $Y$ are the corresponding gauge structure constants; $g_s$, $g=e/{sin\theta_W}$ and $g'=e/{cos\theta_W}$ are the strong and electroweak gauge couplings; $f_s$, $f$ and $f'$ are parameters determined by the composite dynamics.

In order to compute decay widths of the excited quarks, above lagrangian has been implemented into the CompHEP [25]. Obtained results for branching ratios of decays of the excited quarks into SM quarks and gauge bosons are given in Table I. In these calculations, $\Lambda=m^{*}$ and $f=f'=f_s=1$ have been taken. Also, Fig. 1 shows total and partial decay widths as a function of the excited quark mass. More than 80 percent of excited quarks decay into quark and gluon.

\begin{figure}
\subfigure[]{\includegraphics[width=7.15cm]{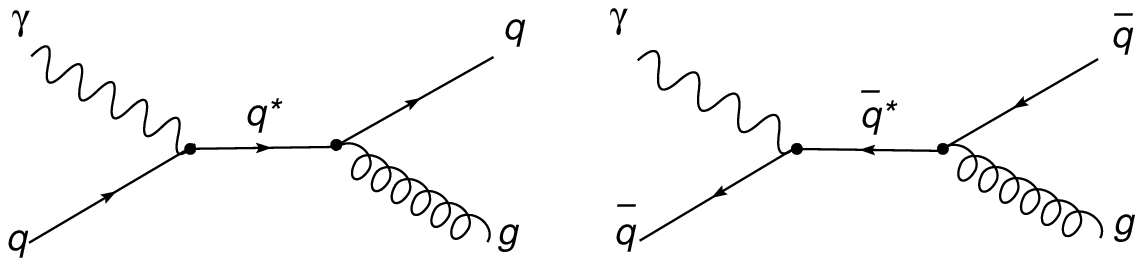}}\quad \quad \quad
\subfigure[]{\includegraphics[width=7.15cm]{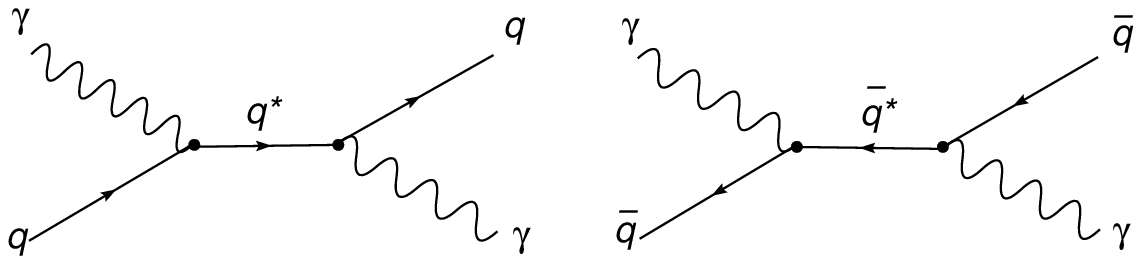}}
\subfigure[]{\includegraphics[width=7.15cm]{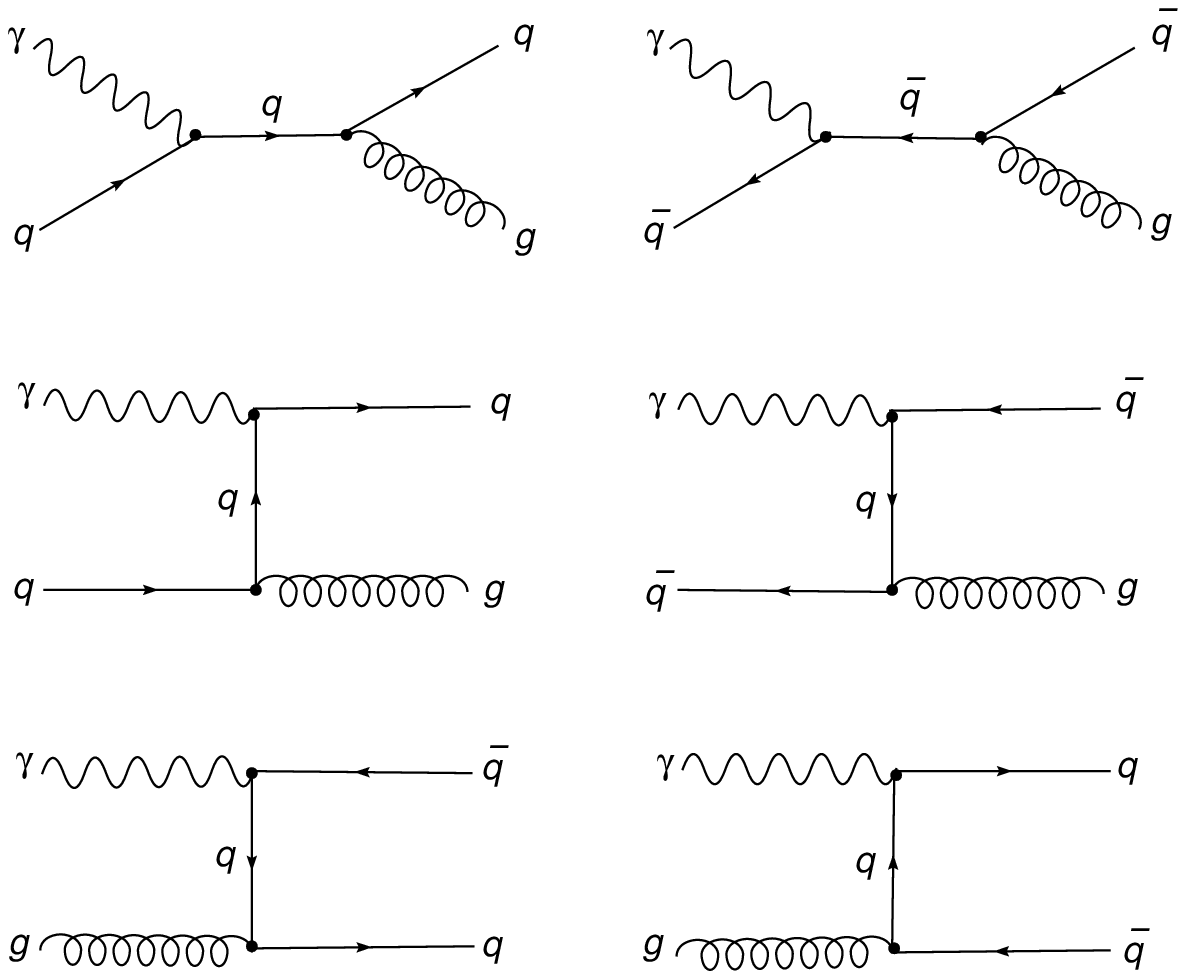}}\quad \quad \quad
\subfigure[]{\includegraphics[width=7.15cm]{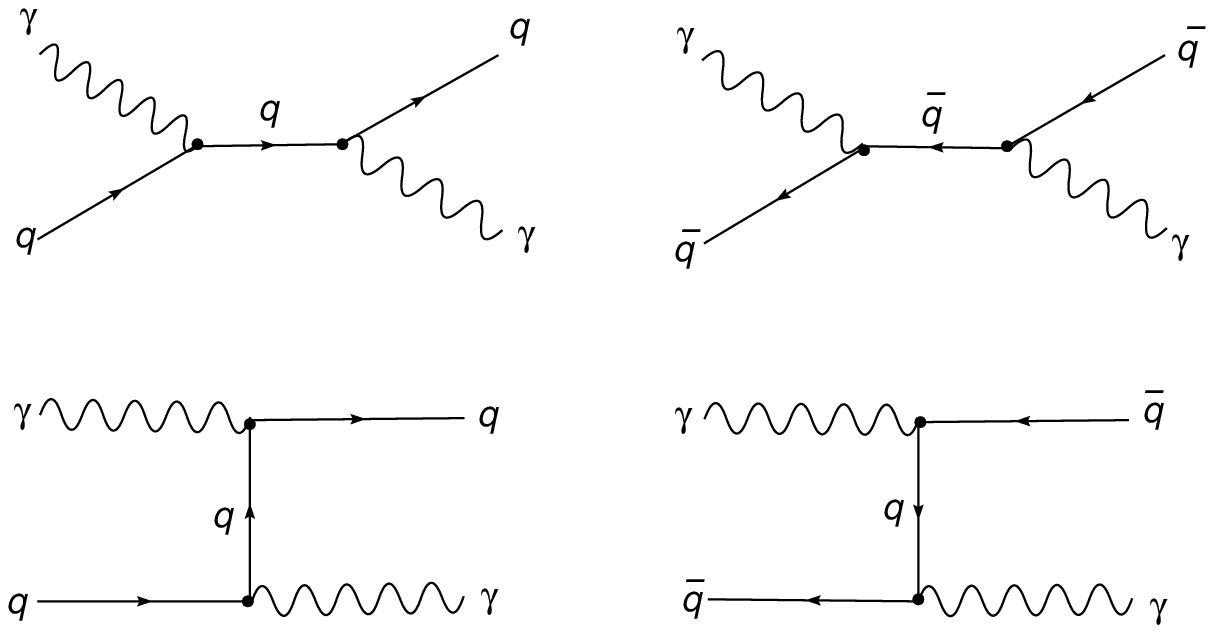}}
\caption{The Feynman diagrams for the excited quark production (a) via the $\gamma q \rightarrow q^{*} \rightarrow g q$ and (b) via the $\gamma q \rightarrow q^{*} \rightarrow \gamma q$; (c) and (d) represent their SM background, respectively. \protect\label{fig3}}
\end{figure}

\begin{table}
\caption{The cross sections of the $m^{*}=1$ TeV excited quark signal ($\Lambda=m^{*}$ and $f=f'=f_s=1$) and corresponding SM background with various $p_T$ (GeV) ranges for QCD Explorer based $\gamma p$ collider. All values are at the unit of pb.}
\begin{ruledtabular}
\begin{tabular}{cccccc}
\multicolumn{2}{c}{Process} &  no $p_T$ cut & $p_T > 100$ GeV & $p_T > 200$ GeV & $p_T > m^*/3$  \\
\colrule
& \textbf{Signal} & 	 & 	&  	 &  	 \\
		&	$u^*$ & $4.15\times 10^{-2}$ & $4.15\times 10^{-2}$ &  $4.13\times 10^{-2}$ &  $3.86\times 10^{-2}$ \\
		&	$d^*$ & $1.93\times 10^{-3}$ & $1.93\times 10^{-3}$ &  $1.90\times 10^{-3}$ &  $1.64 \times10^{-3}$ \\

$\gamma p\rightarrow j j X$		&\textbf{SM backg.}	 &  &  &  &  	 \\
		&	$\gamma q\rightarrow g q$ & $1.43\times 10^{5}$ & 20.44 &  1.62 &  $7.47\times 10^{-2}$ \\
		&	$\gamma \bar{q}\rightarrow g \bar{q}$ & $1.40\times 10^{5}$ & 6.47 &  0.15 &  $1.67\times 10^{-3}$ \\
		&	$\gamma g\rightarrow q \bar{q}$	& $1.50\times 10^{6}$ & 30.78 &  0.58 &  $5.74\times 10^{-3}$ \\
		&	All & $1.79\times 10^{6}$ & 57.69 &  2.35 &  $8.21\times 10^{-2}$ \\
\colrule
& \textbf{Signal} & 	 & 	&  	 &  	 \\
		&	$u^*$ & $1.15\times 10^{-3}$ & $1.15\times 10^{-3}$ &  $1.14\times 10^{-3}$ &  $1.08\times 10^{-3}$ \\
		&	$d^*$ & $1.33\times 10^{-5}$ & $1.33\times 10^{-5}$ &  $1.31\times 10^{-5}$ &  $1.14\times 10^{-5}$ \\
 
$\gamma p\rightarrow \gamma j X$		&\textbf{SM backg.}	 & 	&  & 	& \\
		&	$\gamma q\rightarrow \gamma q$ & $5.26\times 10^{2}$ & 0.50 &  $4.14\times 10^{-2}$ &  $1.94\times 10^{-3}$ \\
		&	$\gamma \bar{q}\rightarrow \gamma \bar{q}$  & $5.02 \times10^{2}$ & 0.14 &  $3.36 \times10^{-3}$ &  $3.70\times 10^{-5}$ \\
		&	All & $1.03\times 10^{3}$ & 0.64 &  $4.47\times 10^{-2}$ &  $1.98\times 10^{-3}$ \\

\end{tabular}
\end{ruledtabular}
\end{table}

\begin{table}
\caption{The cross sections of the $m^{*}=1$ TeV excited quark signal ($\Lambda=m^{*}$ and $f=f'=f_s=1$) and corresponding SM background with various $p_T$ (GeV) ranges for Energy Frontier $ep$ based $\gamma p$ collider. All values are at the unit of pb.}
\begin{ruledtabular}
\begin{tabular}{cccccc}
\multicolumn{2}{c}{Process} &  no $p_T$ cut & $p_T > 100$ GeV & $p_T > 200$ GeV & $p_T > \frac{m^*}{3}$  \\
\colrule
& \textbf{Signal} & 	 & 	&  	 &  	 \\
		&	$u^*$ & 8.96 & 8.96 & 8.60 &  7.16 \\
		&	$d^*$ & 1.38 & 1.38 & 1.32 &  1.08 \\
		
$\gamma p\rightarrow j j X$		&\textbf{SM backg.}	 &  &  &  &  	 \\
		&	$\gamma q\rightarrow g q$ & $2.73\times 10^{4}$ & 62.20 &  8.30 &  1.74 \\
		&	$\gamma \bar{q}\rightarrow g \bar{q}$ & $2.67\times 10^{4}$ & 42.72 &  3.27 &  0.35 \\
		&	$\gamma g\rightarrow q \bar{q}$	& $2.85\times 10^{5}$ & 305.48 &  19.69 &  1.82 \\
		&	All & $3.39\times 10^{5}$ & 410.40 &  31.26 &  3.90 \\
\colrule
& \textbf{Signal} & 	 & 	&  	 &  	 \\
		&	$u^*$ & $2.48\times 10^{-1}$ & $2.48\times 10^{-1}$ &  $2.38\times 10^{-1}$ & $1.95\times 10^{-1}$ \\
		&	$d^*$ & $9.55\times 10^{-3}$ & $9.55\times 10^{-3}$ &  $9.13\times 10^{-3}$ & $7.52\times 10^{-3}$ \\
		
$\gamma p\rightarrow \gamma j X$		&\textbf{SM backg.}	 & 	&  &  	& \\
		&	$\gamma q\rightarrow \gamma q$ & $2.21\times 10^{2}$ & 1.48 &  0.20 &  $4.36\times 10^{-2}$ \\
		&	$\gamma \bar{q}\rightarrow \gamma \bar{q}$  & $2.15\times 10^{2}$ & 0.99 &  0.05 &  $7.63\times 10^{-3}$ \\
		&	All & $4.36\times 10^{2}$ & 2.47 &  0.25 &  $5.12\times 10^{-2}$ \\

\end{tabular}
\end{ruledtabular}
\end{table}     

\section{The Excited Quark Production at $\gamma p$ Colliders}

The production of the first generation excited quarks at $\gamma p$ colliders takes place through the photonic excitation process of $q \gamma \rightarrow q^{*}$. The calculated total production cross sections for $u^*$ and $d^*$ are presented for the QCD Explorer based $\gamma p$ collider ($\sqrt{s_{max}}$ = 1.27 TeV) and the Energy Frontier $\gamma p$ collider ($\sqrt{s_{max}}$ = 3.41 TeV) in Fig. 2. In numerical calculations, CTEQ6L1 parton distribution functions are used [26]. As it can be seen, $u^*$ production is about an order higher than $d^*$. The reasons for this are the number of the corresponding quarks in proton and their respective charges. In this study only $\gamma q \rightarrow q^{*} \rightarrow g q$ and $\gamma q \rightarrow q^{*} \rightarrow \gamma q$ processes are considered as signatures (at parton level) of the excited quarks at $\gamma p$ colliders. These signatures will be observed as dijet and photon-jet at the detector, respectively.

\begin{table}
\caption{The signal and background cross sections of the excited quark $jj$ channel and corresponding SS for QCD Explorer based $\gamma p$ collider.}
\begin{ruledtabular}
\begin{tabular}{ccccc}
$m^{*}$ & \multicolumn{2}{c} {$\gamma p \rightarrow u^{*} X \rightarrow jj X$($\gamma p \rightarrow d^{*} X \rightarrow jj X$)} & \multicolumn{2}{c} {$SS$} \\
\cline{2-3}\cline{4-5}
(GeV) & $\sigma_{S}$ (pb) & $\sigma_{B}$ (pb) & $\textsl{L}_{int}=0.6$ fb$^{-1}$ & $\textsl{L}_{int}=6$ fb$^{-1}$ \\ 
\colrule

600 &  6.310  (0.700) & 2.58 &  96.23 (10.67) &  304.30 (33.76) \\
650 &  3.900  (0.390) & 1.64 &  74.60 (7.46) &  235.90 (23.59) \\
700 &  2.350  (0.210) & 1.08 &  55.39 (4.95) &  175.16 (15.65) \\
750 &  1.350  (0.110) & 0.71 &  39.24 (3.20) &  124.10 (10.11) \\
800 &  0.750  (0.056) & 0.45 &  27.39 (2.04) &  86.60 (6.47) \\
850 &  0.404  (0.027) & 0.31 &  17.77 (1.19) &  56.21 (3.76) \\
900 &  0.201  (0.011) & 0.20 &  11.01 (0.60) &  34.81 (1.91) \\
950 &  0.092  (0.004) & 0.12 &  6.51 (0.28) &  20.57 (0.89) \\
1000 & 0.040  (0.002) & 0.082 &  3.42 (0.17) &  10.82 (0.54) \\
1050 & 0.0151  (0.0006) & 0.052 &  1.62 (0.06) &  5.13 (0.20) \\
1100 & 0.0050  (0.0002) & 0.032 &  0.68 (0.03) &  2.17 (0.09) \\
\end{tabular}
\end{ruledtabular}
\end{table}

\begin{figure}
\subfigure[]{\includegraphics[width=8.15cm]{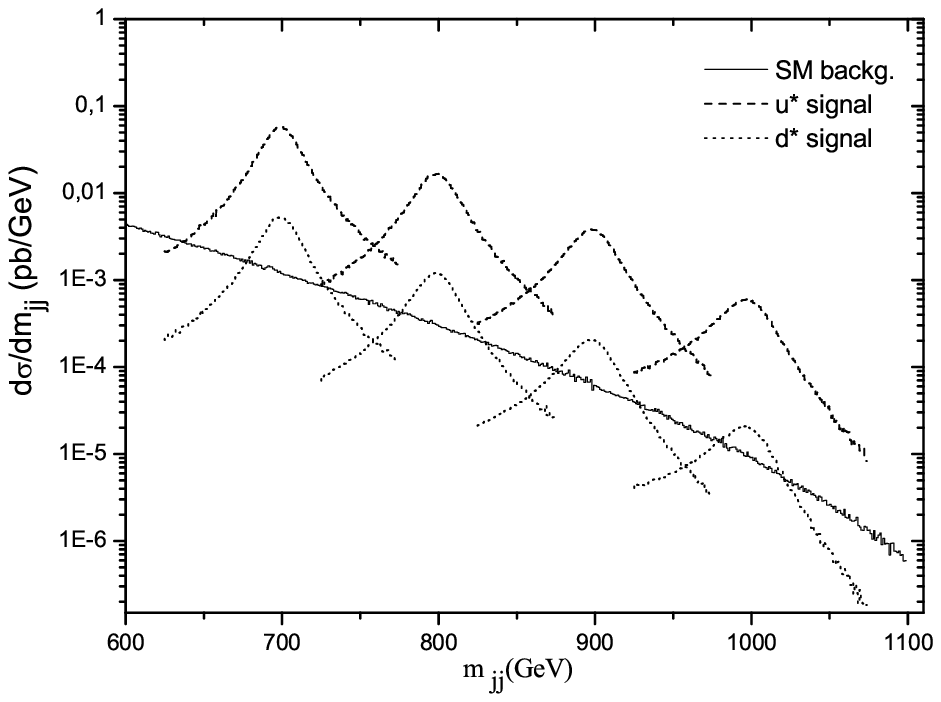}}
\subfigure[]{\includegraphics[width=8.15cm]{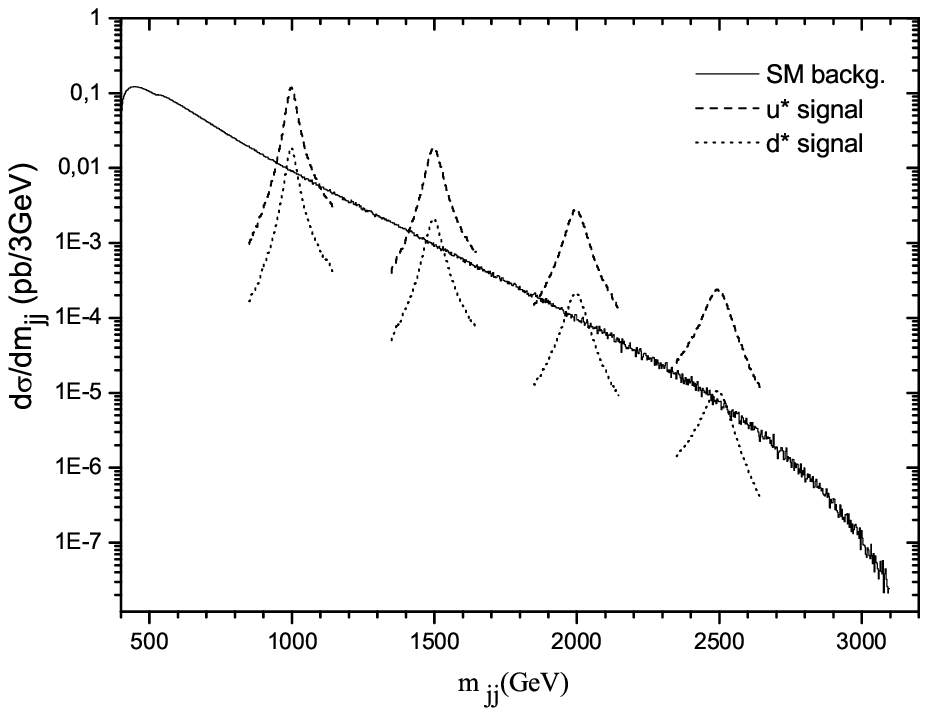}}
\subfigure[]{\includegraphics[width=8.15cm]{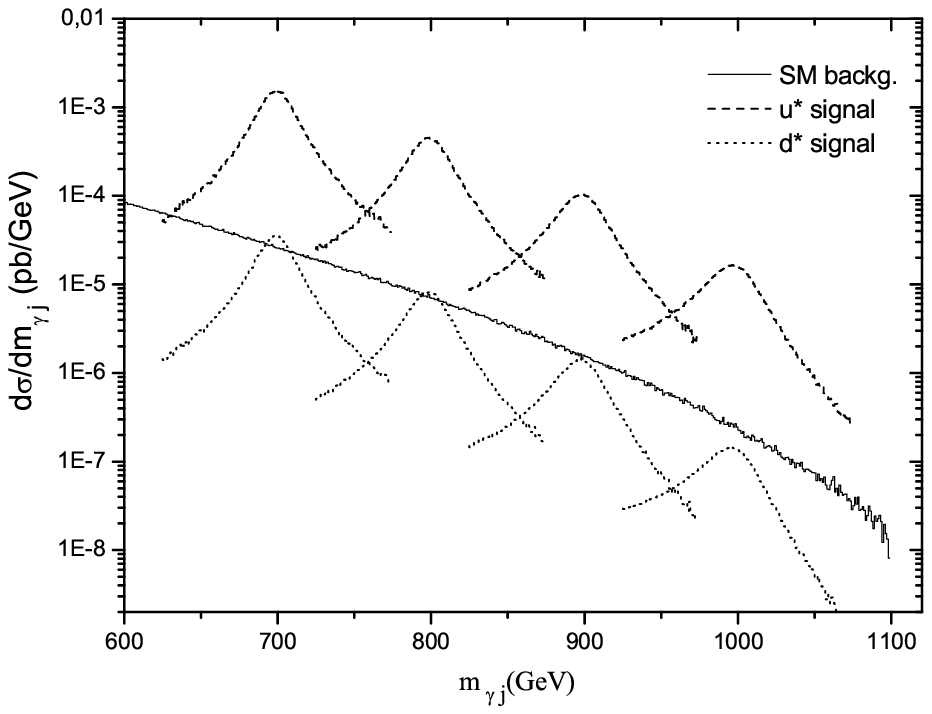}}
\subfigure[]{\includegraphics[width=8.15cm]{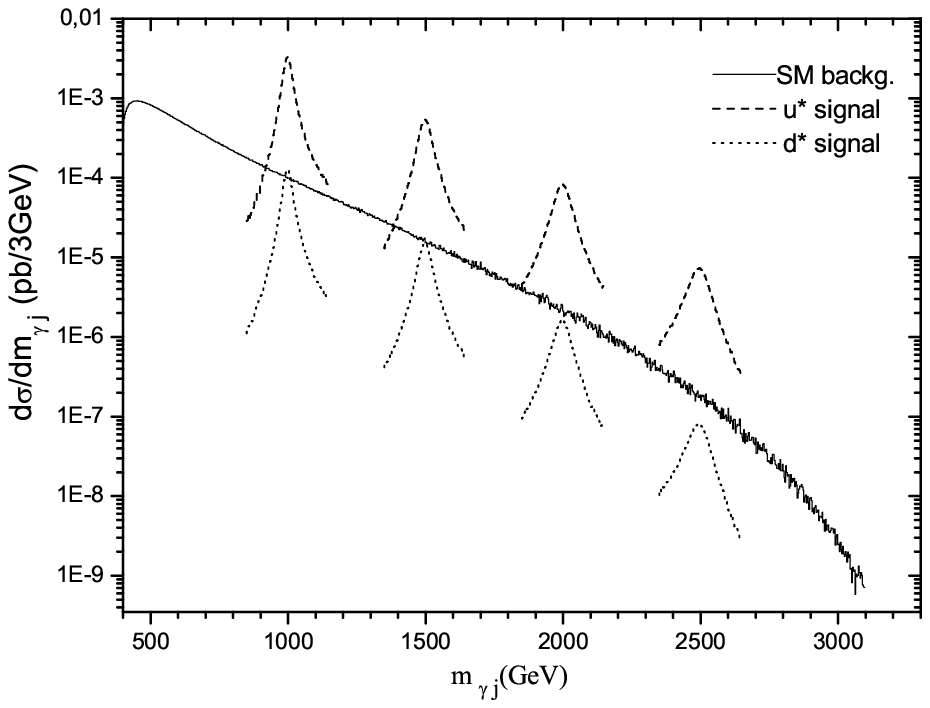}}
\caption{Invariant mass distributions for the excited quark signal (with the scale $\Lambda=m^{*}$ and couplings $f=f'=f_s=1$) and corresponding SM background at $\gamma p$ collider (a) $jj$ channel for $m^{*}=700, 800, 900, 1000$ GeV mass values with $\sqrt{s}$ = 1.27 TeV and (b) for $m^{*}=1000, 1500, 2000, 2500$ GeV mass values with $\sqrt{s}$ = 3.41 TeV; (c) $\gamma j$ channel with $\sqrt{s}$ = 1.27 TeV and (d) $\sqrt{s}$ = 3.41 TeV.\protect\label{fig4}}
\end{figure}

The main tree level Feynman diagrams for the excited quark production at $\gamma p$ collider with subsequent $gq$ and $\gamma q$ decay channels and their SM background processes are shown in Fig. 3. Since the largest contributions to production cross section of the excited quark come from $s$-channel subprocesses, $t$-channel subprocesses are omitted. The second diagrams in Fig. 3a and 3b originate from sea quarks at proton. All the SM processes ($s$- and $t$-channels) allowed in $\gamma p$ collisions yielding two jets (Fig. 3c) and photon-jet (Fig. 3d) were considered as background. For these background processes, both valance and sea quark contributions are considered.

\begin{table}
\caption{The signal and background cross sections of the excited quark $jj$ channel and corresponding SS for Energy Frontier $ep$ based $\gamma p$ collider.}
\begin{ruledtabular}
\begin{tabular}{ccccc}
$m^{*}$ & \multicolumn{2}{c} {$\gamma p \rightarrow u^{*} X \rightarrow jj X$($\gamma p \rightarrow d^{*} X \rightarrow jj X$)} & \multicolumn{2}{c} {$SS$} \\
\cline{2-3}\cline{4-5}
(GeV) & $\sigma_{S}$ (pb) & $\sigma_{B}$ (pb) & $\textsl{L}_{int}=60$ pb$^{-1}$ & $\textsl{L}_{int}=600$ pb$^{-1}$ \\ 
\colrule

600 &  33.60  (6.34) & 33.73 &  44.8 (8.5) &  141.7 (26.7) \\
700 &  21.40  (3.86) & 18.71 &  38.3 (6.9) &  121.2 (21.9) \\
800 &  14.28  (2.46) & 10.27 &  34.5 (5.9) &  109.1 (18.8) \\
900 &  9.98  (1.62) & 6.20 &  31.0 (5.0) &  98.2 (15.9) \\
1000 &  7.16  (1.08) & 3.90 &  28.1 (4.2) &  88.8 (13.4) \\
1100 &  5.19  (0.74) & 2.50 &  25.4 (3.6) &  80.4 (11.5) \\
1200 &  3.81  (0.51) & 1.68 &  22.8 (3.0) &  72.0 (9.6) \\
1300 &  2.81  (0.36) & 1.13 &  20.5 (2.6) &  64.8 (8.3) \\
1400 &  2.07  (0.25) & 0.78 &  18.2 (2.2) &  57.4 (6.9) \\
1500 &  1.53  (0.17) & 0.55 &  16.0 (1.8) &  50.5 (5.6) \\
1600 &  1.12  (0.12) & 0.39 &  13.9 (1.5) &  43.9 (4.7) \\
1700 &  0.81  (0.079) & 0.28 &  11.9 (1.2) &  37.5 (3.7) \\
1800 & 0.59  (0.053) & 0.20 &  10.2 (0.92) &  32.3 (2.9) \\
1900 &  0.41  (0.035) & 0.14 &  8.5 (0.72) &  26.8 (2.3) \\
2000 & 0.29  (0.022) & 0.10 &  7.1 (0.54) &  22.5 (1.7) \\
2100 & 0.19  (0.014) & 0.077 &  5.3 (0.39) &  16.8 (1.2) \\
2200 & 0.13  (0.0085) & 0.057 &  4.2 (0.28) &  13.3 (0.87) \\
2300 & 0.084  (0.0049) & 0.040 &  3.3 (0.19) &  10.3 (0.60) \\
2400 & 0.052  (0.0027) & 0.030 &  2.3 (0.12) &  7.4 (0.38) \\
2500 & 0.031  (0.0014) & 0.022 &  1.6 (0.07) &  5.1 (0.23) \\
2600 & 0.018  (0.00072) & 0.016 &  1.1 (0.044) &  3.5 (0.14) \\
2700 & 0.0094  (0.00033) & 0.011 &  0.69 (0.024) &  2.2 (0.077) \\
\end{tabular}
\end{ruledtabular}
\end{table}

\begin{figure}
\subfigure[]{\includegraphics[width=8.15cm]{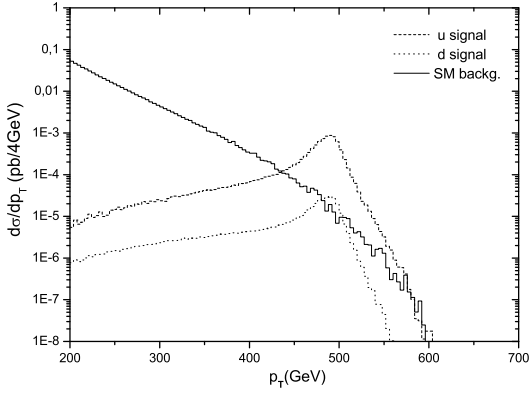}}
\subfigure[]{\includegraphics[width=8.15cm]{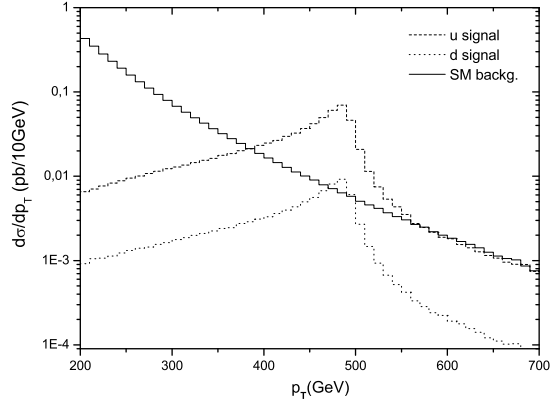}}
\subfigure[]{\includegraphics[width=8.15cm]{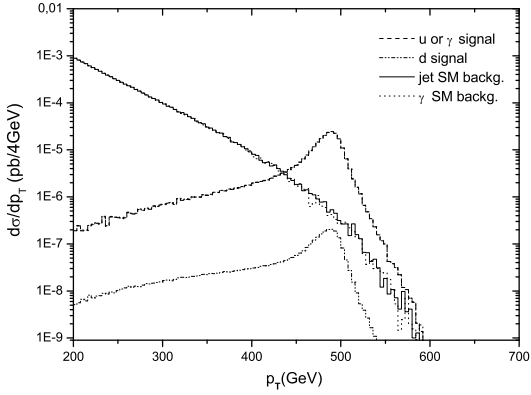}}
\subfigure[]{\includegraphics[width=8.15cm]{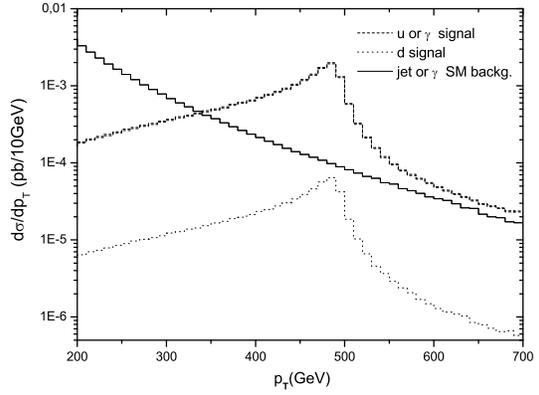}}
\caption{Transverse momentum distributions for the excited quark signal (with the scale $\Lambda=m^{*}=1$ TeV and couplings $f=f'=f_s=1$) and corresponding SM background at $\gamma p$ collider (a) $jj$ channel with $\sqrt{s}$ = 1.27 TeV and (b) $\sqrt{s}$ = 3.41 TeV; (c) $\gamma j$ channel with $\sqrt{s}$ = 1.27 TeV and (d) $\sqrt{s}$ = 3.41 TeV . \protect\label{fig5}}
\end{figure}

\begin{table}
\caption{The signal and background cross sections of the excited quark $\gamma j$ channel and corresponding SS for QCD Explorer based $\gamma p$ collider.}
\begin{ruledtabular}
\begin{tabular}{ccccc}
$m^{*}$ & \multicolumn{2}{c} {$\gamma p \rightarrow u^{*} X \rightarrow \gamma j X$($\gamma p \rightarrow d^{*} X \rightarrow \gamma j X$)} & \multicolumn{2}{c} {$SS$} \\
\cline{2-3}\cline{4-5}
(GeV) & $\sigma_{S}$ (pb) & $\sigma_{B}$ (pb) & $\textsl{L}_{int}=0.6$ fb$^{-1}$ & $\textsl{L}_{int}=6$ fb$^{-1}$ \\ 
\colrule

600 &  0.17  (0.0045) & 0.045 &  19.6 (0.52) &  62.1 (1.6) \\
650 &  0.10  (0.0026) & 0.031 &  13.9 (0.36) &  44.0 (1.1) \\
700 &  0.062  (0.0014) & 0.022 &  10.2 (0.23) &  32.4 (0.73) \\
750 &  0.037  (0.00074) & 0.014 &  7.7 (0.15) &  24.2 (0.48) \\
800 &  0.020  (0.00037) & 0.0099 &  4.9 (0.091) &  15.6 (0.29) \\
850 &  0.011  (0.00017) & 0.0069 &  3.2 (0.050) &  10.3 (0.16) \\
900 &  0.0054  (0.000078) & 0.0045 &  2.0 (0.028) &  6.2 (0.090) \\
950 &  0.0025  (0.000031) & 0.0030 &  1.1 (0.014) &  3.5 (0.044) \\
1000 & 0.0011  (0.000011) & 0.0020 &  0.60 (0.0060) &  1.9 (0.019) \\
\end{tabular}
\end{ruledtabular}
\end{table}

\begin{figure}
\subfigure[]{\includegraphics[width=8.15cm]{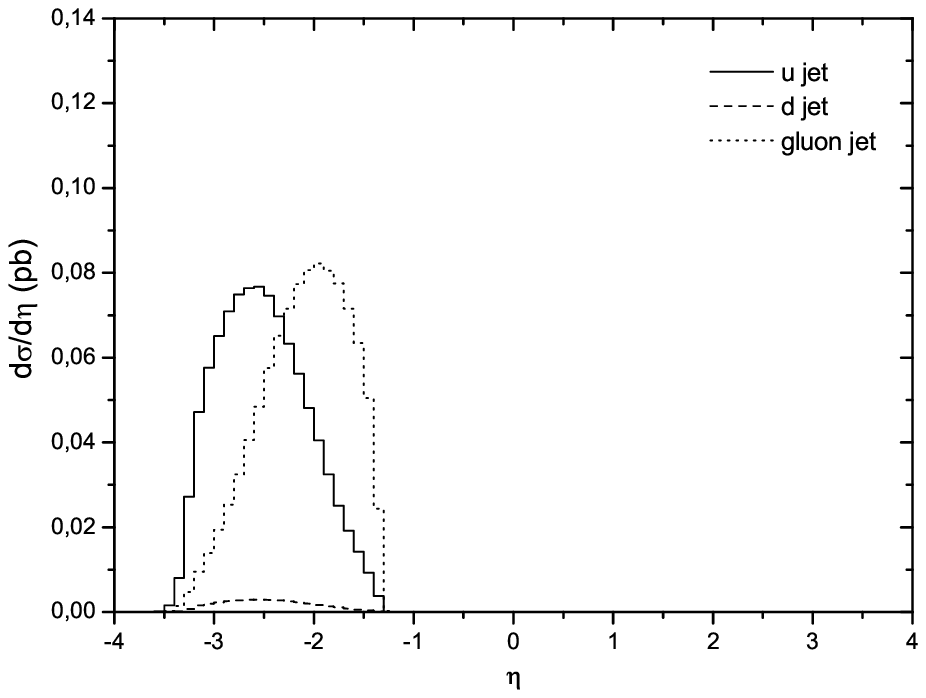}}
\subfigure[]{\includegraphics[width=8.15cm]{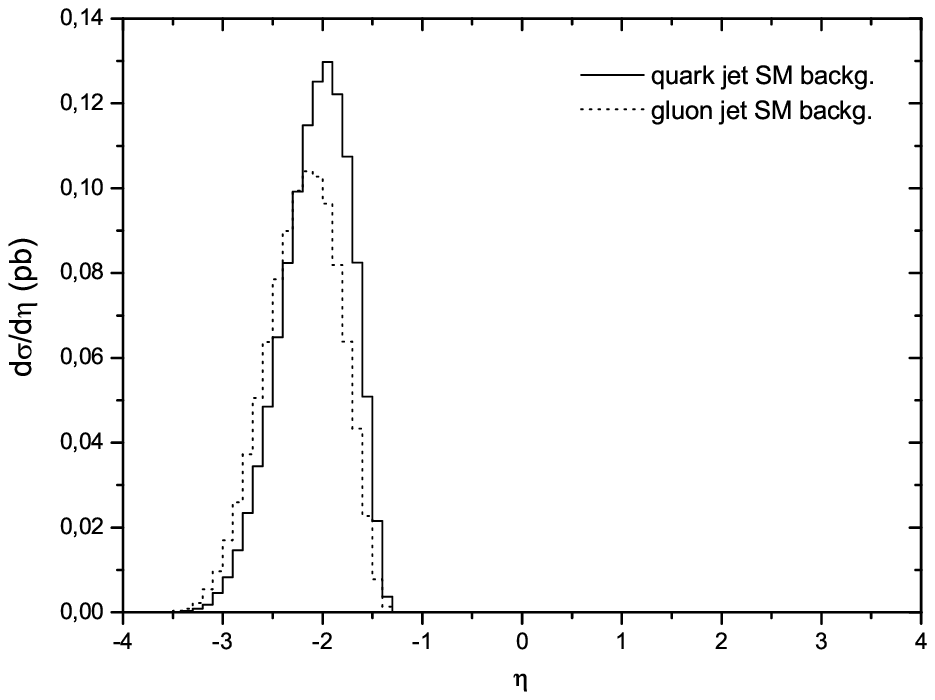}}
\subfigure[]{\includegraphics[width=8.15cm]{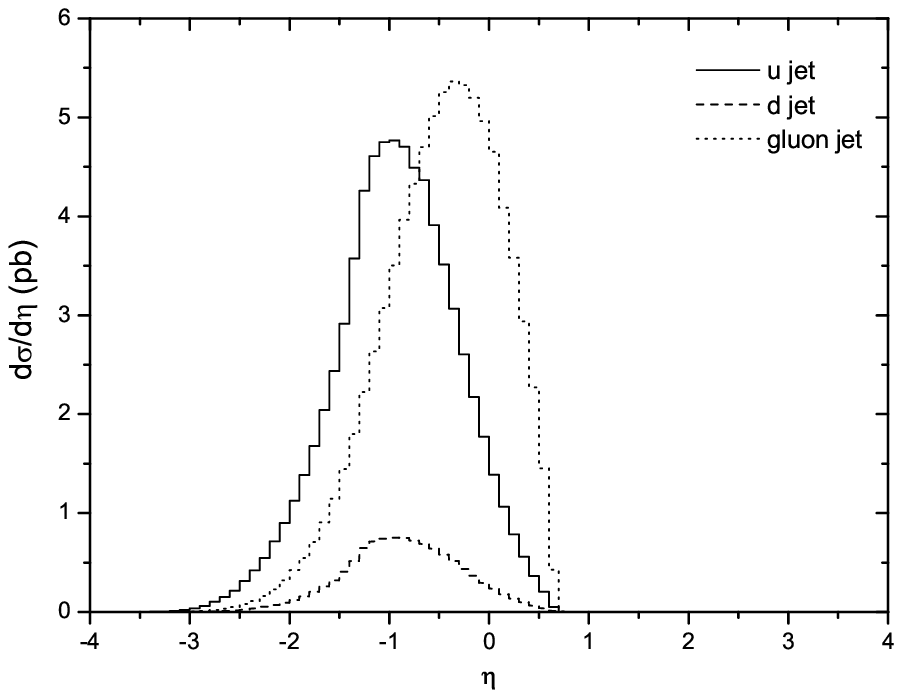}}
\subfigure[]{\includegraphics[width=8.15cm]{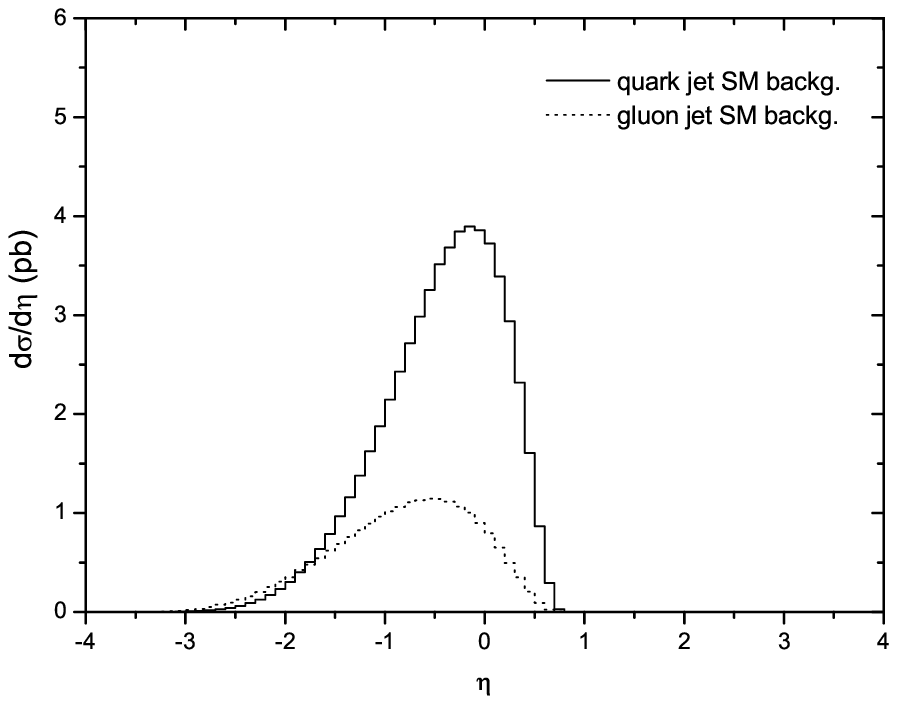}}
\caption{Pseudorapidity distributions for the excited quark signal in the $jj$ channel (with the scale $\Lambda=m^{*}=1$ TeV and couplings $f=f'=f_s=1$) and corresponding SM background at $\gamma p$ collider (a), (b) with $\sqrt{s}$ = 1.27 TeV and (c), (d) with $\sqrt{s}$ = 3.41 TeV. \protect\label{fig6}}
\end{figure}

\begin{table}
\caption{The signal and background cross sections of the excited quark $\gamma j$ channel and corresponding SS for Energy Frontier $ep$ based $\gamma p$ collider.}
\begin{ruledtabular}
\begin{tabular}{ccccc}
$m^{*}$ & \multicolumn{2}{c} {$\gamma p \rightarrow u^{*} X \rightarrow \gamma j X$($\gamma p \rightarrow d^{*} X \rightarrow \gamma j X$)} & \multicolumn{2}{c} {$SS$} \\
\cline{2-3}\cline{4-5}
(GeV) & $\sigma_{S}$ (pb) & $\sigma_{B}$ (pb) & $\textsl{L}_{int}=60$ pb$^{-1}$ & $\textsl{L}_{int}=600$ pb$^{-1}$ \\ 
\colrule

600 &  0.78  (0.041) & 0.28 &  11.4 (0.60) &  36.1 (1.9) \\
700 &  0.57  (0.025) & 0.17 &  10.7 (0.47) &  33.9 (1.5) \\
800 &  0.38  (0.016) & 0.11 &  8.9 (0.37) &  28.1 (1.2) \\
900 &  0.27  (0.011) & 0.074 &  7.7 (0.31) &  24.3 (0.99) \\
1000 &  0.19  (0.0075) & 0.051 &  6.5 (0.26) &  20.6 (0.81) \\
1100 &  0.14  (0.0052) & 0.036 &  5.7 (0.21) &  18.1 (0.67) \\
1200 &  0.10  (0.0036) & 0.026 &  4.8 (0.17) &  15.2 (0.55) \\
1300 &  0.080  (0.0025) & 0.019 &  4.5 (0.14) &  14.2 (0.44) \\
1400 &  0.060  (0.0017) & 0.014 &  3.9 (0.11) &  12.4 (0.35) \\
1500 &  0.045  (0.0012) & 0.010 &  3.5 (0.093) &  11.0 (0.29) \\
1600 &  0.033  (0.00085) & 0.0078 &  2.9 (0.075) &  9.2 (0.24) \\
1700 &  0.024  (0.00058) & 0.0058 &  2.4 (0.059) &  7.7 (0.19) \\
1800 & 0.017  (0.00039) & 0.0044 &  2.0 (0.046) &  6.3 (0.14) \\
1900 &  0.012  (0.00025) & 0.0032 &  1.6 (0.034) &  5.2 (0.11) \\
2000 & 0.0086  (0.00017) & 0.0024 &  1.4 (0.027) &  4.3 (0.085) \\
2100 & 0.0059  (0.00010) & 0.0018 &  1.1 (0.018) &  3.4 (0.058) \\
2200 & 0.0039  (0.000064) & 0.0014 &  0.81 (0.013) &  2.6 (0.042) \\
\end{tabular}
\end{ruledtabular}
\end{table}

For both the dijet and photon-jet processes, the pseudorapidity of jets is taken $\left| \eta_j \right|<2.5$; the pseudorapidity of photon is taken $\left| \eta_{\gamma} \right|<2.5$; separation between jets is taken $\Delta R_{j j}>0.7$ and separation between photon and jet is taken $\Delta R_{j \gamma}>0.4$ as detector limitations. As seen from Tables II and III one should note that with above detector cuts the SM background cross section originate from the processes: $\gamma q\rightarrow g q$ (where $q=u,d,c,s$) contributes \%8, $\gamma \bar{q}\rightarrow g \bar{q}$ (where $q=u,d,c,s$) contributes \%8, and $\gamma g\rightarrow q \bar{q}$ (where $q=u,d,c,s,t,b$) contributes \%84 in $\gamma p\rightarrow j j X$ process; $\gamma q\rightarrow \gamma q$ (where $q=u,d,c,s$) contributes \%51, and $\gamma \bar{q}\rightarrow \gamma \bar{q}$ (where $q=u,d,c,s$) contibutes \%49 in $\gamma p\rightarrow \gamma j X$ process. In order to extract the excited quark signal and to suppress the SM background, we impose transverse momentum cut to jets also. Various $p_T$ cuts are applied to both signal and background processes. As seen from Tables II and III, the most of the SM backgrounds reduced with $p_T$ cut; optimal cut value for transverse momentum seems to be about one third of the excited quark mass. Consequently, in the numerical calculations transverse momenta of jets are taken bigger than $m^*/3$ to suppress the background. One should not that, transverse momentum cut is more efficient for QCD Explorer. For example, while the dijet SM background is decreased by order of $10^5$ for Energy Frontier, it is reduced with more than $10^7$ factor for QCD Explorer. By applying above detector and $p_T$ cuts the computed signal and background cross sections for the $\gamma p \rightarrow q^{*} X \rightarrow jj X$ process are given in Tables IV and V for $\sqrt{s_{max}}$ = 1.27 TeV and $\sqrt{s_{max}}$ = 3.41 TeV options, respectively. Similar calculations for the $\gamma p \rightarrow q^{*} X \rightarrow \gamma j X$ process are presented in Tables VI and VII. In the last two columns of these tables, the statistical significance (SS) values of the excited quark signal (evaluated from $SS=(\sigma_{S}/\sqrt{\sigma_B})\sqrt{L_{int}}$, where $L_{int}$ is the integrated luminosity of the collider) are given.           

One way to observe excited quark is to look at cross section with respect to the jet-jet invariant mass for the $\gamma p \rightarrow q^{*} X \rightarrow jj X$ channel. The invariant mass spectra for the signal and the SM background are given in Fig. 4. It is drawn for sample values of excited quark masses of 700, 800, 900 and 1000 GeV for the QCD Explorer based $\gamma p$ collider and 1000, 1500, 2000 and 2500 GeV for the Energy Frontier $\gamma p$ collider in Fig. 4a and 4b, respectively. As can be seen from the figure, the expected signal for $u^*$ is well over the background, while $d^*$ signal is somewhat comparable with background, in the region of the peak. Similarly Fig. 4c and 4d show the distribution of the reconstructed invariant mass of the photon-jet for the $\gamma p \rightarrow q^{*} X \rightarrow \gamma j X$ channel for QCD Explorer and Energy Frontier based $\gamma p$ colliders, respectively. Since $d^*$ signal becomes well below background with the increase of the photon-jet invariant mass, it will be difficult to observe excited $d$ quark.

\begin{figure}
\subfigure[]{\includegraphics[width=8.15cm]{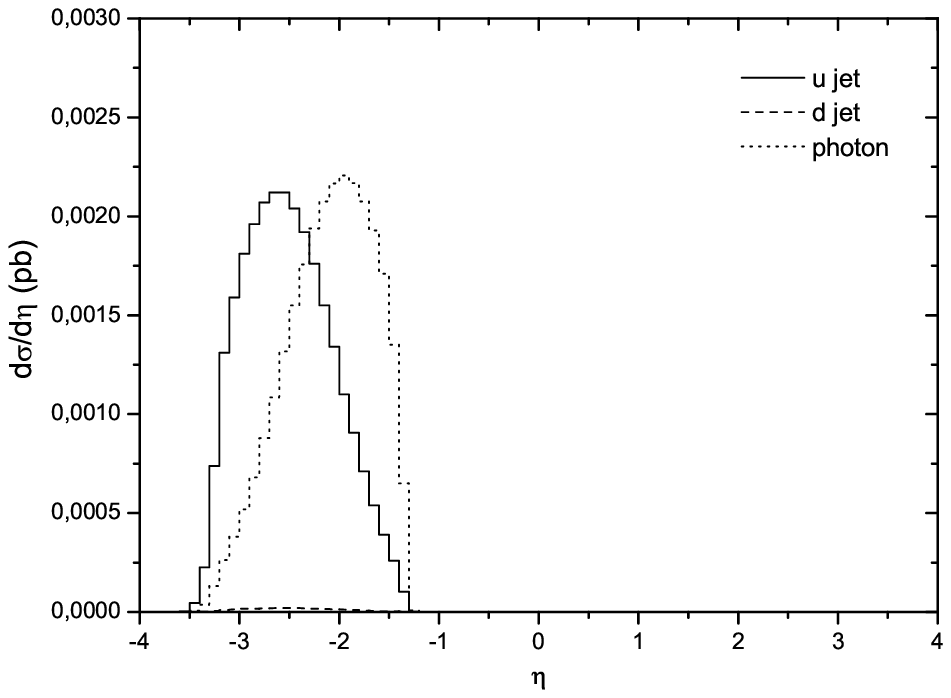}}
\subfigure[]{\includegraphics[width=8.15cm]{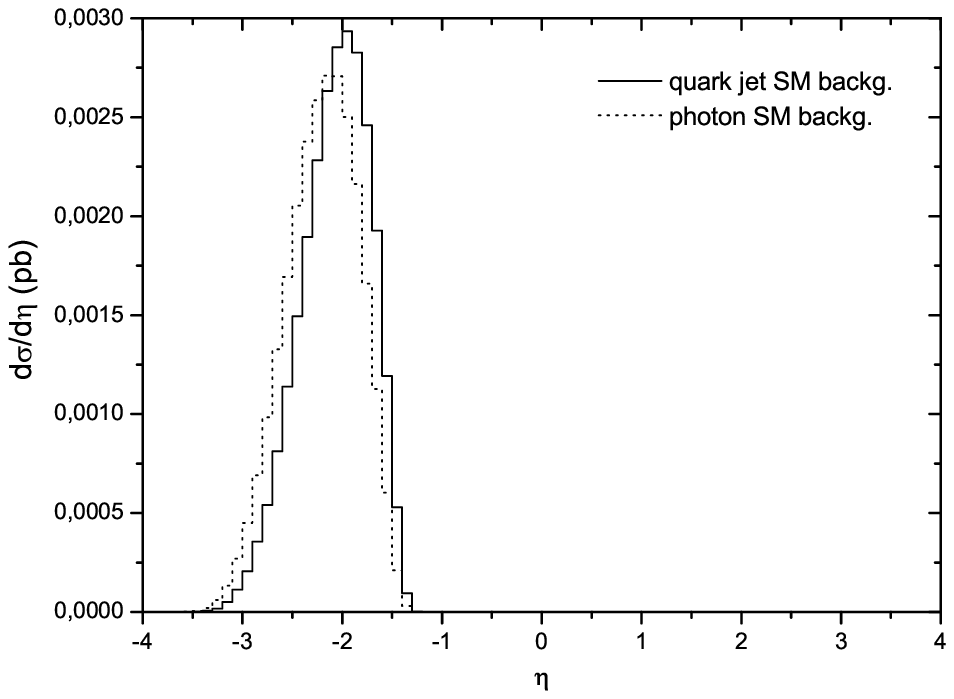}}
\subfigure[]{\includegraphics[width=8.15cm]{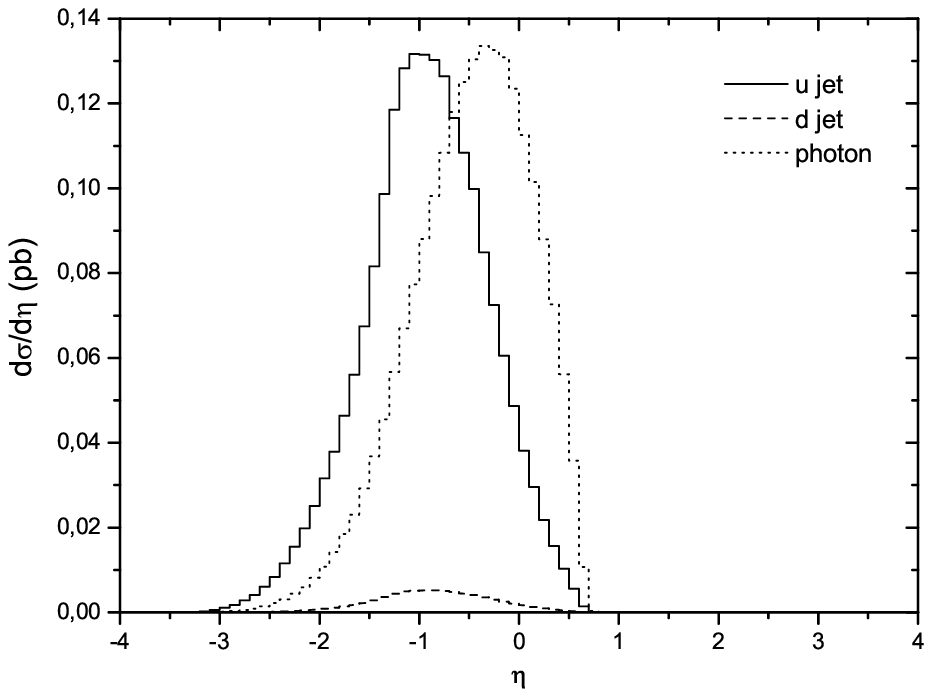}}
\subfigure[]{\includegraphics[width=8.15cm]{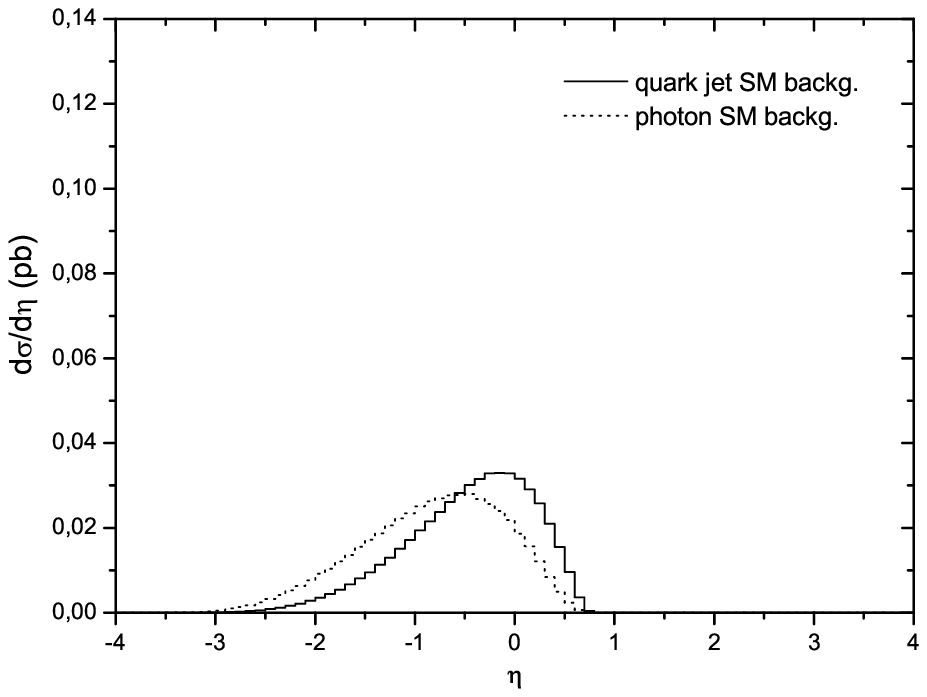}}
\caption{Pseudorapidity distributions for the excited quark signal in the $\gamma j$ channel (with the scale $\Lambda=m^{*}=1$ TeV and couplings $f=f'=f_s=1$) and corresponding SM background at $\gamma p$ collider (a), (b) with $\sqrt{s}$ = 1.27 TeV and (c), (d) with $\sqrt{s}$ = 3.41 TeV. \protect\label{fig7}}
\end{figure}  

The transverse momentum distributions for the SM backgrounds as well as the signal of the $jj$ and $\gamma j$ channels are given in Fig. 5(a,b) and 5(c,d), respectively, for the excited quark mass of 1 TeV in the cases $\Lambda=m^{*}$ and $f=f'=f_s=1$. As can be seen from figures, the transverse momentum peaks at the half of the excited quark mass; photons and jets originating from $u$ quarks have high transverse momenta compare with SM backgrounds. Hence these figures could help us to determine optimal $p_T$ cut value and eliminate most of the corresponding background.

\begin{figure}
\subfigure[]{\includegraphics[width=8.15cm]{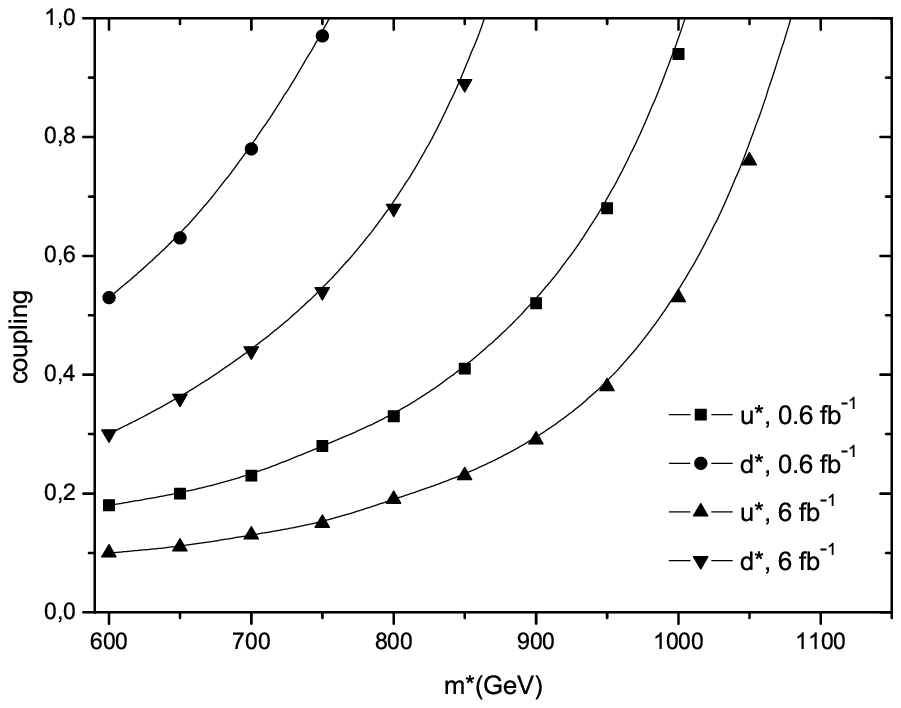}}
\subfigure[]{\includegraphics[width=8.15cm]{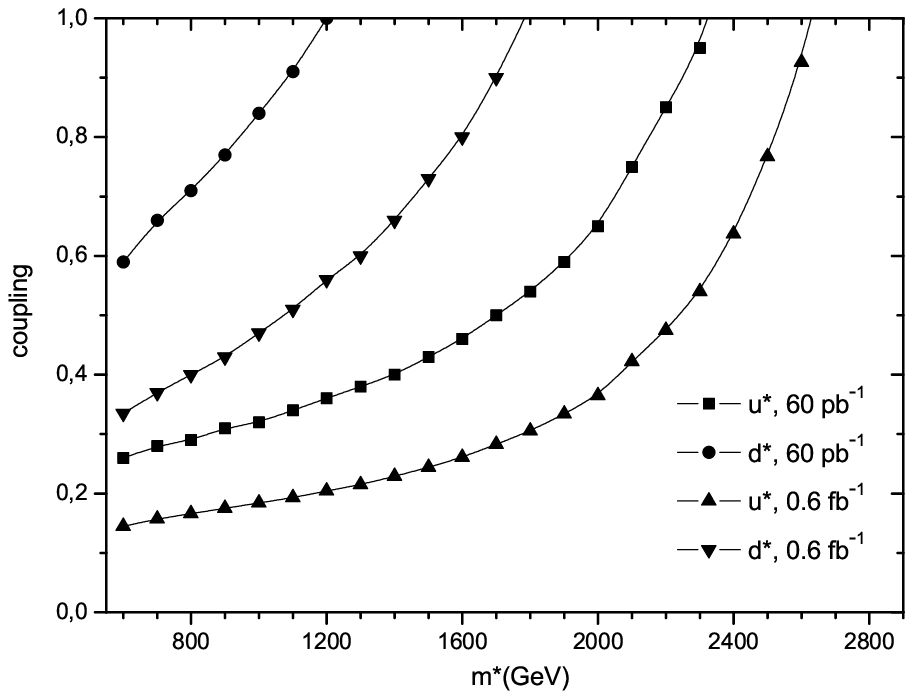}}
\caption{Observation reach at $3\sigma$ for coupling and excited quark mass ($jj$ channel) at $\gamma p$ collider with (a) $\sqrt{s}$ = 1.27 TeV and (b) $\sqrt{s}$ = 3.41 TeV. \protect\label{fig8}}
\end{figure}

Pseudorapidity, $\eta$, is a spatial coordinate describing the angle of a particle relative to the beam axis. It is defined as

\begin{equation}
\eta=- ln\left[tan\left(\frac{\theta}{2}\right)\right]
\end{equation}
\\
where $\theta$ is the polar angle relative to the $\gamma$ beam axis in this paper. The signal and SM background distributions for jets in dijet channel with respect to the pseudorapidity are shown in Fig. 6. As seen from the figure, pseudorapidity distributions of jets resulting from u and d quarks are mostly negative because of momentum boost on the direction of proton beam. Since QCD Explorer is more asymmetrical compared to Energy Frontier, the pseudorapidity distribution of QCD Explorer is more negative shifted than ones for Energy Frontier. As conferred from Fig. 6a and 6b, the pseudorapidity of QCD Explorer spans from -1.5 to -3.1 corresponding to polar angles of 25 degrees to 5 degrees relative to proton beam axis. In this paper, $\left| \eta_j \right|<2.5$ is taken as detector cut on pseudorapidity which is similar to ATLAS detector cut. However, half of the signal is left out with this choice. Therefore one needs to design a special asymmetric detector for QCD Explorer based $\gamma p$ collider. Energy Frontier $\gamma p$ collider has $ -2.5 \leq \eta_j \leq 0.5 $ for the dijet channel as seen from Fig. 6c and 6d. ATLAS type detector is suitable for this collider. Similar pseudorapidity ranges are determined for photon jet channel in both QCD Explorer and Energy Frontier $\gamma p$ colliders (Fig. 7).

\begin{figure}
\subfigure[]{\includegraphics[width=8.15cm]{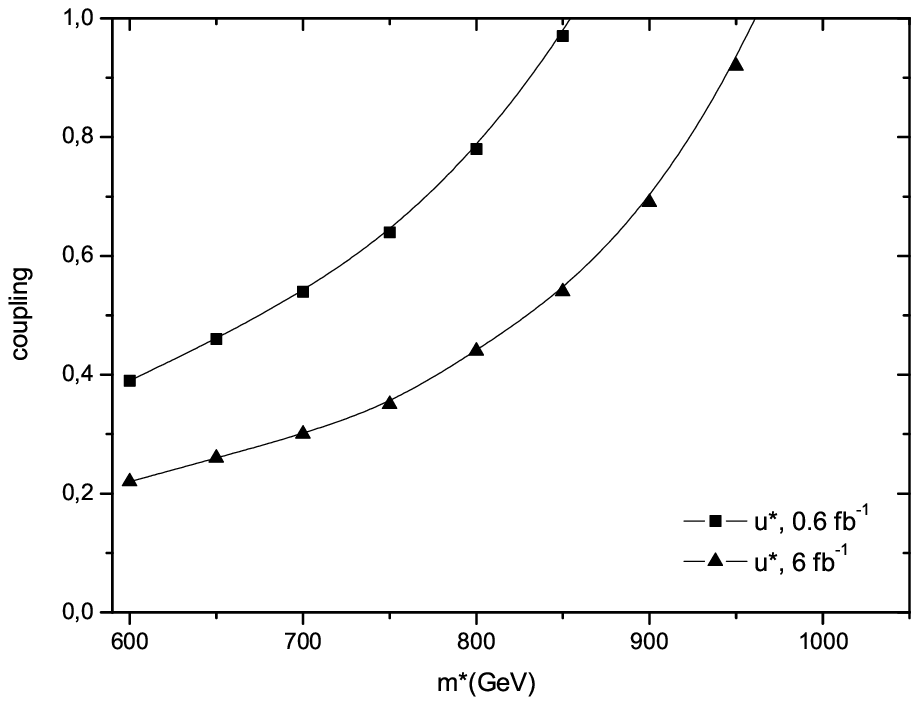}}
\subfigure[]{\includegraphics[width=8.15cm]{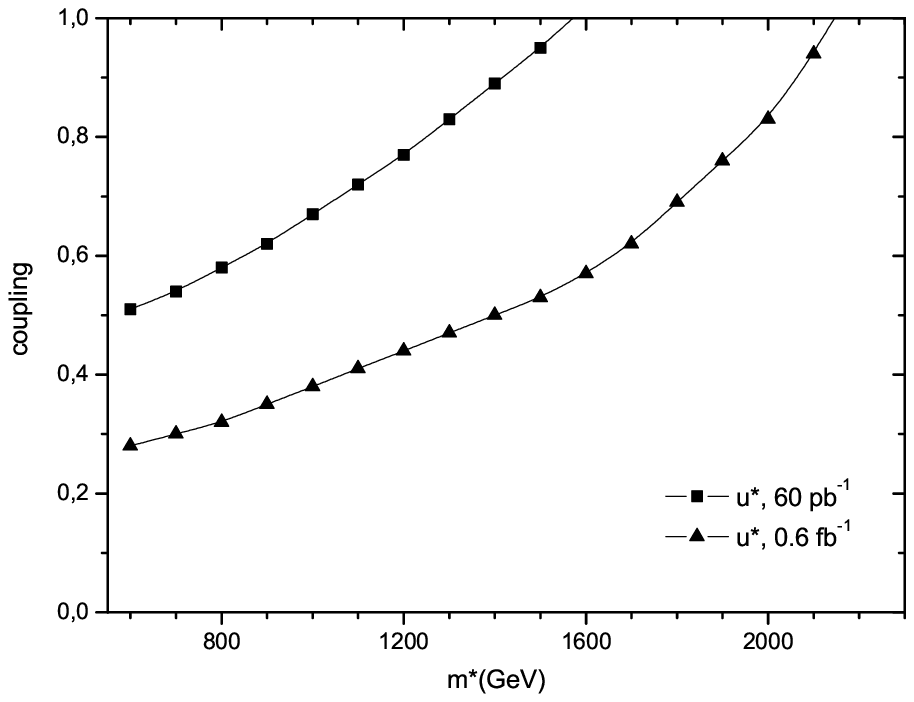}}
\caption{Observation reach at $3\sigma$ for coupling and excited quark mass ($\gamma j$ channel) at $\gamma p$ collider with (a) $\sqrt{s}$ = 1.27 TeV and (b) $\sqrt{s}$ = 3.41 TeV. \protect\label{fig9}}
\end{figure}

Achievable mass and coupling values for the excited quark are established by requiring $SS\geq3$ as an observation criterion. For the process $\gamma p \rightarrow q^{*} X \rightarrow jj X$ it is seen from Tables IV and V, that QCD Explorer will cover masses of $u^{*}$ excited quark up to 1000 GeV (1080 GeV) with ${L}_{int}=0.6$ fb$^{-1}$ (6 fb$^{-1}$) for $\Lambda=m^{*}$ and $f=f'=f_s=1$, whereas Energy Frontier $\gamma p$ collider will extend the mass region up to $m_{u^{*}}=2320$ GeV (2620 GeV) with ${L}_{int}=60$ pb$^{-1}$ (600 pb$^{-1}$). Similarly, QCD Explorer will reach $m_{d^{*}}=755$ GeV (870 GeV) with ${L}_{int}=0.6$ fb$^{-1}$ (6 fb$^{-1}$). Corresponding limit for Energy Frontier is 1200 GeV (1790 GeV) with ${L}_{int}=60$ pb$^{-1}$ (600 pb$^{-1}$). Observation reach of compositeness parameters $f_s$, $f$ and $f^\prime $ as a function of the excited quark mass for process under consideration is presented in Fig. 8 for different $\gamma p$ colliders. As seen from the figure, it will be possible to probe compositeness parameters down to 0.1 depending on the mass.  

For the process $\gamma p \rightarrow q^{*} X \rightarrow \gamma j X$ Tables VI and VII show that, excited up quark will be observed up to $m_{u^{*}}=855$ GeV (965 GeV) at QCD Explorer with ${L}_{int}=0.6$ fb$^{-1}$ (6 fb$^{-1}$), whereas these limits becomes 1570 GeV (2140 GeV) for Energy Frontier with ${L}_{int}=60$ pb$^{-1}$ (600 pb$^{-1}$). Figure 9 shows observation reach of compositeness parameters $f_s$, $f$ and $f^\prime $ as a function of the excited up quark mass. Observation of $d^{*}$ excited quark is not possible at this channel with given collider parameters (see Tables VI and VII). 

\section{Conclusion}

$\gamma p$ colliders are complementary to hadron colliders like LHC. While LHC is the best place for discovery of excited quarks in a very wide range of masses, $\gamma p$ colliders such as ones that based on QCD Explorer and Energy Frontier $ep$ collider are essential for detailed studies of properties of the excited quark interactions. Concerning the criterion $SS\geq3$ and taking $\Lambda=m^{*}$ and $f=f'=f_s=1$, the excited quark can be observed up to 2.6 TeV at future $\gamma p$ colliders. In this paper, only $\gamma p \rightarrow q^{*} X \rightarrow jj X$ and $\gamma p \rightarrow q^{*} X \rightarrow \gamma j X$ channels are considered as signatures. It is determined that the former channel is more promising for both $u^{*}$ and $d^{*}$ than latter one. For example, if $m_{q^{*}}=850$ GeV, the numbers of produced jet jet (photon jet) events are 2424 (66) for $u^{*}$ and 162 (not observable) for $d^{*}$ at $\sqrt{s_{max}}$ = 1.27 TeV with ${L}_{int}=6$ fb$^{-1}$. If $m_{q^{*}}=1700$ GeV, the numbers of produced jet jet (photon jet) events are 486 (14) for $u^{*}$ and 47 (not observable) for $d^{*}$ at $\sqrt{s_{max}}$ = 3.41 TeV with ${L}_{int}=600$ pb$^{-1}$. 

Also, this study shows that QCD Explorer based $\gamma p$ collider requires a forward detector designed for it. Meanwhile one can use a detector similar to ATLAS for Energy Frontier $\gamma p$ collider.            

\begin{acknowledgments}

I would like to thank A. K. Ciftci and S. Sultansoy for many helpful conversations and discussions, and acknowledge for support from the Scientific and Technical Research Council (TUBITAK) BIDEB-2218 grant. This work was also supported in part by the State Planning Organization (DPT) under grant no DPT-2006K-120470 and in part by the Turkish Atomic Energy Authority (TAEA) under grant no VII-B.04.DPT.1.05.
 
\end{acknowledgments}

\end{document}